\documentclass[conference,a4paper]{IEEEtran}
\usepackage[left=1.43cm,right=1.43cm,top=1.8cm,bottom=4.21cm]{geometry}

\usepackage[algo2e]{algorithm2e} 
\usepackage{amsmath,amssymb,amsmath,amsfonts}
\usepackage{bm}
\usepackage{bbm}
\usepackage{graphicx}
\usepackage{cite}
\usepackage{epstopdf}
\usepackage{gensymb}
\usepackage{color}
\usepackage{lipsum}
\usepackage{float}
\usepackage{epstopdf}
\usepackage[mathcal]{eucal}
\usepackage{subfiles}
\usepackage[T1]{fontenc}
\usepackage{siunitx}
\usepackage{tabulary}
\usepackage{epsfig,graphics,graphicx,subfig,amssymb,amstext,amsmath,algorithm,algorithmic,wrapfig,multirow}
\usepackage{array}
\usepackage{adjustbox}
\usepackage[dvipsnames]{xcolor}
\usepackage{cite}
\usepackage{times}
\usepackage{placeins}
\usepackage{textcomp}
\usepackage[font=small]{caption}
\usepackage{flushend}
\usepackage{color, colortbl}
\usepackage{tabularx}
\usepackage{bm}
\usepackage{enumerate}
\usepackage{tikz}
\usepackage{pgfplots}
\usepackage{epstopdf}
\usetikzlibrary{shapes,arrows}
\usepackage[utf8]{inputenc}
\usepackage{mathtools}
\usepackage[utf8]{inputenc}
\usepackage{xcolor}
\usepackage{tikz}
\usetikzlibrary{chains,arrows,calc,positioning}
\usepackage{amssymb}% http://ctan.org/pkg/amssymb
\usepackage{pifont}% http://ctan.org/pkg/pifont
\usepackage{soul}
\usepackage{breqn} % to split equations using dmath env
\usepackage{booktabs} % nice rules in tables
\usepackage{multirow}
\usepackage{enumitem}
\usepackage{tabulary}
\usepackage[normalem]{ulem}
\usepackage{dblfloatfix}
\usepackage{makecell}
\usepackage{lipsum}
\usepackage{amsthm}
\usepackage{comment}
\usepackage{filecontents}
\usepackage{hyperref}

\usepackage{listings}
\newtheoremstyle{mystyle}%                % Name
  {}%                                     % Space above
  {}%                                     % Space below
  {\itshape}%                             % Body font
  {}%                                     % Indent amount
  {\bfseries}%                            % Theorem head font
  {.}%                                    % Punctuation after theorem head
  { }%                                    % Space after theorem head, ' ', or \newline
  {}%                                     % Theorem head spec (can be left empty, meaning `normal')
\theoremstyle{mystyle}

\newlength \figwidth
\definecolor{bittersweet}{rgb}{1.0, 0.44, 0.37}
\definecolor{glaucous}{rgb}{0.38, 0.51, 0.71}
\definecolor{gainsboro}{rgb}{0.86, 0.86, 0.86}
\definecolor{babyblueeyes}{rgb}{0.63, 0.79, 0.95}
\definecolor{silver}{rgb}{0.75, 0.75, 0.75}
\definecolor{neoncarrot}{rgb}{1.0, 0.64, 0.26}
\definecolor{Gray}{gray}{0.9}
\definecolor{LightCyan}{rgb}{0.88,1,1}
\definecolor{BackgroundLightBlue}{rgb}{0.97,0.97,1}
\definecolor{BackgroundGray}{gray}{0.98}

\newcommand{\red}[1]{{\textcolor[rgb]{1,0,0}{#1}}}

\makeatletter

 \let\oldforeign@language\foreign@language
 \DeclareRobustCommand{\foreign@language}[1]{%
   \lowercase{\oldforeign@language{#1}}}

\def\nb0{{\mathbf{0}}}
\def\nb1{{\mathbf{1}}}

\makeatother

\IEEEoverridecommandlockouts
\begin{document}

% This code is to reduce the list of authors by using et. al:
\bstctlcite{IEEEexample:BSTcontrol}

\title{TSpec-LLM: An Open-source Dataset for\\LLM Understanding of 3GPP Specifications}

\author{
\IEEEauthorblockN{Rasoul Nikbakht$^{\flat}$, Mohamed Benzaghta$^{\star}$, and Giovanni Geraci$^{\dagger\,\star}$ \vspace{0.1cm}
}
\IEEEauthorblockA{$^{\flat}$\emph{Centre Tecnologic de Telecomunicacions de Catalunya (CTTC), Barcelona, Spain}}
\vspace{0.05cm}
\IEEEauthorblockA{$^{\star}$\emph{Universitat Pompeu Fabra (UPF), Barcelona, Spain} \quad\quad\quad $^{\dagger}$\emph{Telefónica Research, Barcelona, Spain}}
%\IEEEauthorblockA{$^{\dagger}$\emph{Telefónica Research, Barcelona, Spain}}

\thanks{This work was in part supported by the Spanish Research Agency through grants PID2021-123999OB-I00, CEX2021-001195-M, and CNS2023-145384, by the UPF-Fractus Chair, and by the Spanish Ministry of Economic Affairs and Digital Transformation and the European Union NextGenerationEU.}
\thanks{The TSpec-LLM dataset presented in this paper, along with the questionnaire and prompts are available open-source \cite{telcospec}.
}
% \vspace{-8mm}
}

\maketitle
%\IEEEpeerreviewmaketitle

\begin{abstract}

Understanding telecom standards involves sorting through numerous technical documents, such as those produced by the 3rd Generation Partnership Project (3GPP), which is time-consuming and labor-intensive. While large language models (LLMs) can assist with the extensive 3GPP knowledge base, an inclusive dataset is crucial for their effective pre-training and fine-tuning. In this paper, we introduce \textit{TSpec-LLM}, an open-source comprehensive dataset covering all 3GPP documents from Release 8 to Release 19 (1999--2023). To evaluate its efficacy, we first select a representative sample of 3GPP documents, create corresponding technical questions, and assess the baseline performance of various LLMs. We then incorporate a retrieval-augmented generation (RAG) framework to enhance LLM capabilities by retrieving relevant context from the \textit{TSpec-LLM} dataset. Our evaluation shows that using a naive-RAG framework on \textit{TSpec-LLM} improves the accuracy of GPT-3.5, Gemini 1.0 Pro, and GPT-4 from 44\%, 46\%, and 51\% to 71\%, 75\%, and 72\%, respectively.

\end{abstract}
%\begin{IEEEkeywords}
%\end{IEEEkeywords}

\section{Introduction}

The 3rd Generation Partnership Project (3GPP) is the primary standardization organization in telecommunications, responsible for developing universal standards for 3G, 4G, and 5G since 1999. In developing these standards, 3GPP collaborates with key industry players to produce comprehensive technical specification documents. Keeping up with the extensive content of these documents (and with new releases) is challenging and time-consuming due to the sheer volume of information. 
Engineers and researchers dealing with 3GPP documents can find a compelling solution in large language models (LLMs), which can develop proficiency in understanding the extensive 3GPP standards knowledge base. LLMs, which fall under the umbrella of natural language processing and generative AI (GenAI), have become a focal point of interest within the wireless communications community \cite{bariah2023large, maatouk2023large, bariah2023understanding, maatouk2023teleqna, lin2023pushing, du2023power, tarkoma2023ai, kotaru2023adapting, holm2021bidirectional, shen2023large, karim2023spec5g, koudouridis2024evaluation, bornea2024telco, piovesan2024telecom, shao2024wirelessllm, saad2024artificial}. By utilizing textual documents from standard specifications and research reports, LLMs are anticipated to play a significant role in designing and operating wireless networks \cite{bariah2023large, maatouk2023large}.

Motivated by this, the authors in \cite{holm2021bidirectional} demonstrated a case study for adapting a BERT-like LLM to telecommunications, where the model was fine-tuned to perform question-answer tasks within the telecom domain. In \cite{maatouk2023teleqna}, the authors provided the first benchmark dataset designed to evaluate LLM knowledge in telecommunications. Both studies demonstrated the competence of base LLM models in handling general telecom-related queries; however, these models encounter difficulties with complex questions related to standards. Although these base models are trained on a large quantity of web data, technical specifications related to wireless communication technologies, even though publicly available, are not easily processed by LLMs due to their complex nature, which includes tables, formulas, and figures. This complexity makes it difficult for LLMs to extract relevant information to accurately respond to user queries, thus limiting the effectiveness of state-of-the-art LLMs in generating precise answers \cite{kotaru2023adapting}.

To enhance LLM performance in the telecom domain, the authors in \cite{karim2023spec5g} provided the open-source dataset SPEC5G, containing a sample of specific 3GPP specification documents merged into a single text file. In SPEC5G, the sampled documents undergo a filtering process that removes symbols, Unicode characters, tables, and their content. While this filtering helps streamline the dataset generation, it also omits information within the tables, which is important as they consist of system parameters, configurations, and other technicalities.
Despite its value as an initial effort, SPEC5G contains only a specific subset of documents and not all releases and their corresponding series. This highlights the need for a more comprehensive and well-structured dataset based on all 3GPP specifications, suitable for utilization by LLMs. 
An inclusive dataset is crucial for the pre-training and fine-tuning of LLMs to effectively understand the telecom domain. This enables LLMs to provide engineers and researchers with an assistant model capable of comprehending and organizing 3GPP technical documents without human intervention.

In this paper, we provide an open-source comprehensive dataset based on all releases of 3GPP specifications called \emph{TSpec-LLM}, ideal for research on utilizing LLMs for telecom. To evaluate the dataset's effectiveness, we perform an assessment study by selecting representative samples of 3GPP documents and creating a set of questions based on technical specifications. We then enhance the LLMs' abilities by implementing a retrieval-augmented generation (RAG) framework capable of retrieving relevant context from the TSpec-LLM dataset for inclusion in the LLM processing window. The main contributions and findings of our paper are as follows:

\begin{itemize}
\item
\textbf{TSpec-LLM dataset:}
We provide a well-structured, comprehensive open-source dataset suitable for conducting research using LLMs. TSpec-LLM retains the original content from the tables and formulas within the 3GPP specifications. Furthermore, TSpec-LLM includes a complete collection of all 3GPP documents spanning from Release 8 to Release 19, published between 1999 and 2023, totaling 13.5\,GB with 30,137 documents and 535 million words. Each document within a specific release preserves its original structure and the contents are neither sampled nor consolidated into a single file.
\item 
\textbf{Automated questionnaire generation:} 
We design a questionnaire derived from 3GPP Releases 15--17, specifically series 36 and 38, using a three-step procedure. Initially, we employ a prompt engineering technique to generate technical questions via the GPT-4 API, which categorizes each question by difficulty. Subsequently, an open-source LLM, Mistral 7 × 8B, verifies these difficulty ratings, followed by a final step of human validation.
\item \textbf{Accuracy assessment:} 
Based on TSpec-LLM and the created questionnaire, we assess state-of-the-art LLMs (i.e., GPT-3.5, GPT-4 \cite{openai2023gpt4}, and Gemini Pro 1.0 \cite{team2023gemini}) by illustrating their performance on intricate questions related to standards specifications. The results indicate that current state-of-the-art LLMs still lack a deep understanding of telecom standards specifications. All models achieve similar accuracy, with GPT-4 performing the best at 51\%, followed by Gemini at 46\%, and GPT-3.5 at 44\%.
\item \textbf{Retrieval-augmented generation (RAG):} 
We propose a naive-RAG framework capable of enhancing the performance of LLMs by integrating two key components. The first is a retrieval system that effectively sources relevant information from the TSpec-LLM dataset, identifying the appropriate document within a certain release based on the user's query to ensure precise data retrieval. The second component focuses on generating responses, such as crafting summaries based on the retrieved document sections. This approach improves the LLM's accuracy and equips it to tackle complex, domain-specific questions more effectively without the need for fine-tuning. Our findings show that by applying a RAG framework to the TSpec-LLM dataset, the accuracy of state-of-the-art LLMs can be enhanced to 71\% and above.
\end{itemize}
\section{The TSpec-LLM Dataset}

\subsection{Functionalities and Limitations of LLMs}
\label{llm-explained}

LLMs are neural networks primarily based on the transformer architecture, capable of processing extensive sets of unlabeled textual data to comprehend and generate human-like language \cite{vaswani2017attention}. The self-attention mechanism, central to the transformer architecture, captures statistical patterns and relationships within the data, allowing the model to predict and generate information. In this mechanism, each word within an input sequence assesses all other words, computing attention scores that indicate the importance of each word in relation to the rest. This enables the model to capture long-distance correlations and understand each word's context. Multi-head attention, another crucial element, enhances this by allowing the model to focus on different aspects of the input, capturing diverse patterns and dependencies \cite{wang2020enhancing}. This helps the model understand complex relationships among words and gain a comprehensive understanding of the input.

LLMs undergo extensive pre-training on large volumes of text data to develop an understanding of the inherent statistical properties of the language. This involves adjusting a model's weights to enable it to forecast the next word in a sentence based on the preceding words. Due to the diverse corpora used in training and the large number of model parameters, LLMs can acquire a thorough grasp of grammar, reasoning capabilities, and even comprehend complex language structures, including foundational wireless communication concepts if contained in the pre-training dataset.

Although state-of-the-art LLMs like the GPT series \cite{brown2020language, openai2023gpt4}, Gemini \cite{team2023gemini}, and others demonstrate remarkable language and knowledge mastery, surpassing human levels in several evaluation benchmarks \cite{srivastava2022beyond}, they still exhibit shortcomings when dealing with specific domains or highly specialized queries \cite{kandpal2023large}. For example, when a user's query extends beyond the model’s pre-training data or requires recent information, LLMs may fail to provide accurate answers. Such limitations pose significant challenges when deploying LLM-based applications in real-world production environments.

Enhancing an LLM's usability entails post-processing steps, including the use of retrieval-augmented generation (RAG) and fine-tuning. RAG combines a retrieval system with a generative model to enhance responses by using documents retrieved from a large corpus as a knowledge base. Fine-tuning adjusts the parameters of a pre-trained model on a specific dataset or task to improve its performance \cite{bai2022training}.
Availing of a carefully curated dataset is therefore crucial for effectively utilizing RAG and fine-tuning techniques to enhance LLM performance on telecom-specific tasks. Our open-source TSpec-LLM dataset aims to close this gap by providing a comprehensive dataset that maintains the structure of the underlying 3GPP documents.

\begin{comment}
Fine-tuning guides an LLM to focus on text distributions useful to humans or within specific domains like wireless communication. The RLHF process creates a reward model based on human feedback collected from interactions with models like GPT-4, developed by OpenAI \cite{openai2023gpt4}, and Llama 3 Instruct, developed by Meta \cite{Llama2}. Enhanced through RLHF, a model can generate AI feedback as a substitute for human feedback.
However, RLHF is unstable and requires a separate reward model. Direct Preference Optimization (DPO) \cite{rafailov2023direct} addresses these issues by using the LLM itself as the reward model, relying only on user preferences. Iterative improvement through curated instructions (high-quality tokens guiding the model) is another area of research. Algorithms like Self-Play Fine-Tuning (SPIN) aim to align the LLM's output probability distribution with curated datasets \cite{chen2024self-spin}. This approach outperforms DPO after 1-2 iterations, but its performance is limited by the quality of the underlying curated dataset.
\end{comment}

\subsection{TSpec-LLM Dataset Creation}

The TSpec-LLM dataset contains processed documentation files from the 3GPP standards, converted to markdown (.md) format to facilitate natural language processing applications. It is intended for use by engineers and researchers utilizing LLMs in the telecommunications sector.

The documents were downloaded from the 3GPP website \cite{3gpp2024} using the open-source tool \emph{download3gpp 0.7.0} \cite{download_3gpp2020}, which downloads all documents from all releases and series to a directory. The dataset was then processed using a custom-designed Python script, available open-source \cite{telcospec}. The script uses the command-line version of LibreOffice and processes the files in parallel, significantly speeding up the conversion. This headless conversion process is suitable for server-side operations and batch-processing scripts, resulting in a well-structured and versatile dataset, particularly suited for natural language processing using LLMs in the telecom domain. Unlike the SPEC5G dataset \cite{karim2023spec5g}, which filters out significant content, TSpec-LLM preserves the full content of each document, totaling 535 million words compared to SPEC5G's 134 million words. An example visualization of the TSpec-LLM dataset is given in the Appendix.

\begin{figure}
\centering
\includegraphics[width=\figwidth]{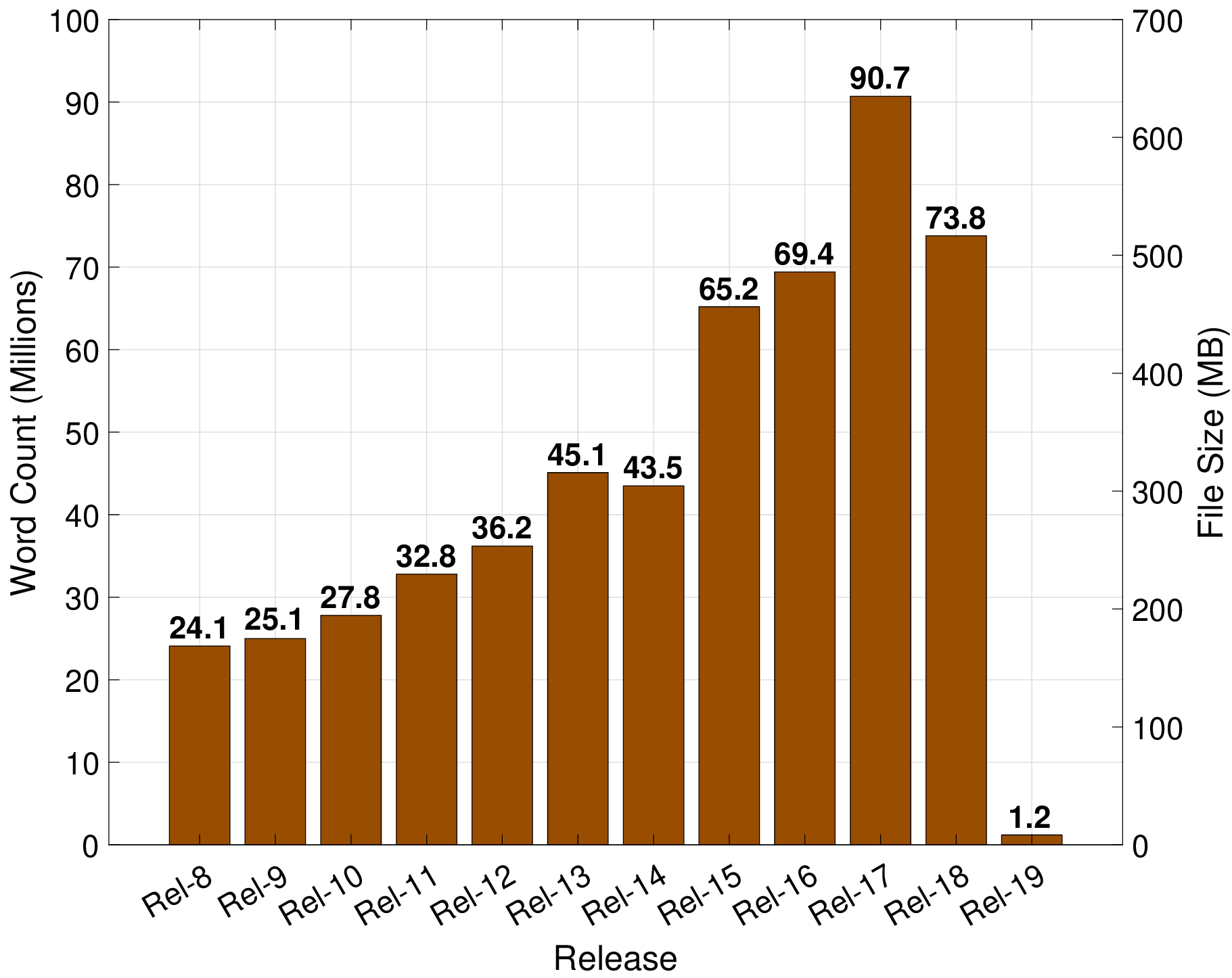}
\caption{Word counts and file sizes for the TSpec-LLM dataset across various 3GPP releases. Data cut-off is December 2023.}
\label{fig:dataset_tokens}
\end{figure}

We used a Python script to analyze the file size of markdown documents in the \texttt{3GPP-clean} directory. This analysis covered all releases and versions of the 3GPP documentation, calculating the total size of .md files. The results were compiled into a report, categorized by version and release, and saved in JSON format. Fig.~\ref{fig:dataset_tokens} reports the total size in megabytes for each release, as well as the total word counts in millions.

%%%%%%%%%%%%%%%%%%%%%%%%%%%%%%%%%%%%
%%%%%%%%%%%%%%%%%%%%%%%%%%%%%%%%%%%%
%\subsection{Insights to the Dataset}

\begin{comment}
\hl{Should this be in the conclusions? Or is it necessary to keep reading?} \red{As the 3GPP documentation contains many formulas and image captions, future versions of the dataset will use an ML-based processing tool to handle mathematical formulas more efficiently} \footnote{\url{https://github.com/transpect/docx2tex} \hl{better in a reference}} \footnote{\url{https://github.com/VikParuchuri/marker} \hl{better in a reference}}.
\end{comment}
%\input{02_LLM_background}
%\input{03_Dataset Creation}
%\input{04_benefit}

\section{Using the TSpec-LLM Dataset for RAG}

Among other applications, the TSpec-LLM dataset can be utilized within the context of RAG. In the following, we introduce the RAG framework, detail how to employ TSpec-LLM for RAG, and report the resulting performance gains.

\subsection{Applying RAG on TSpec-LLM}

\begin{figure}
\centering
\includegraphics[width=.99\figwidth]{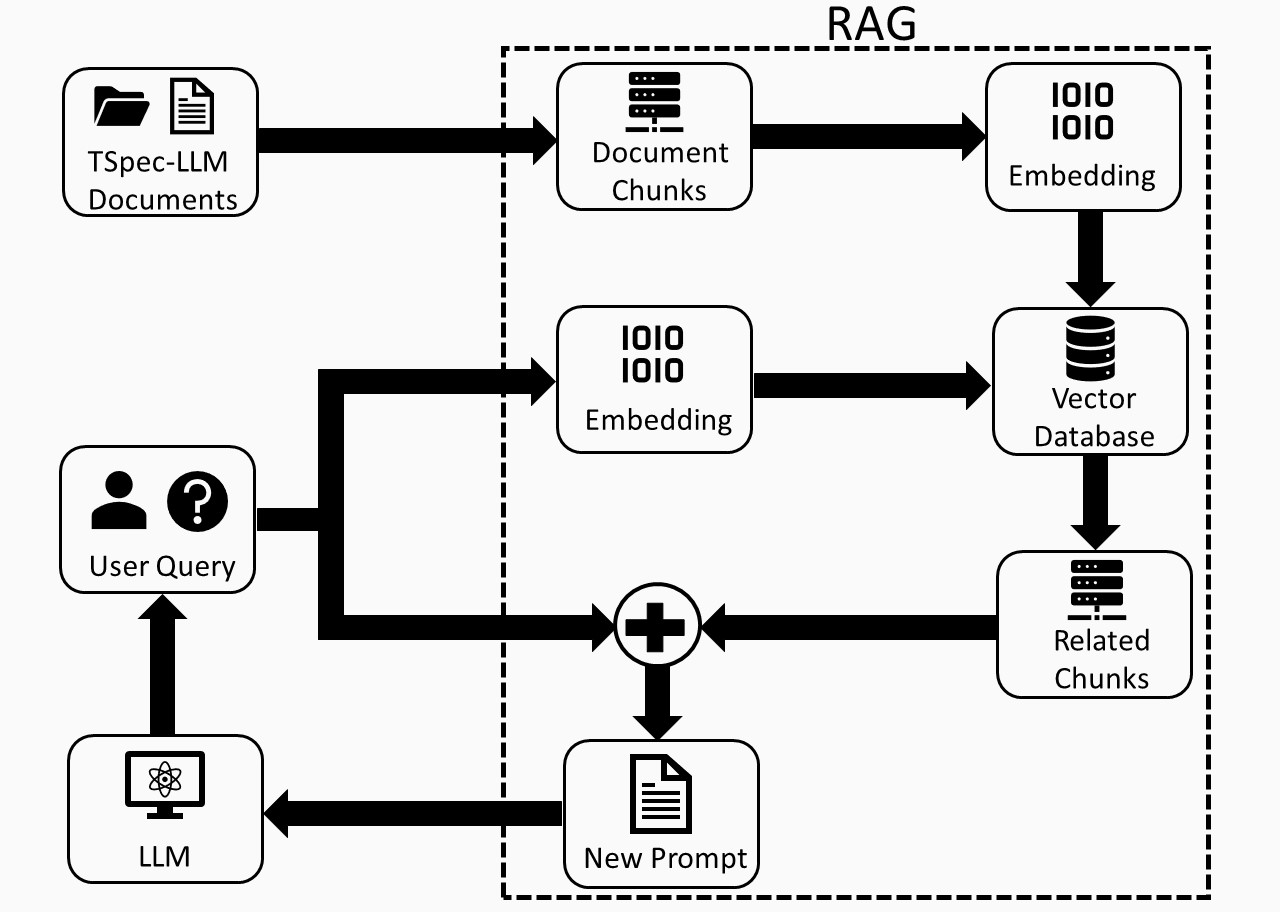}
\caption{Illustration of the naive-RAG paradigm, where documents are divided into chunks and stored in a vector database. User queries are matched with relevant chunks, which are then used to generate a prompt for an LLM to provide a coherent response.}
\label{fig:RAG_framework}
\end{figure}

The RAG framework used consist primarily of an embedding model that assesses the similarity among text segments in the 3GPP specifications. This model transforms text blocks, limited by a window size of $W$ tokens, into vectors within a $D$-dimensional space, converting textual strings to the mathematical space $\textit{R}^D$. For tasks like responding to telecom-related queries, the specialized TSpec-LLM database is segmented into distinct chunks, each smaller than $W$ tokens. These segments are processed offline to generate their respective embeddings. Open-source platforms, such as LlamaIndex \cite{Liu_LlamaIndex_2022} facilitate this process and are compatible with various embedding models. The inference phase of the RAG framework embeds the user's query prompt to identify the vectors in the embedded TSpec-LLM vector database that most closely matches the query. The RAG framework retrieves the top $K$ vectors with the highest similarity, along with the corresponding text segments. Following this, the relevant text to the original user query is used to generate a prompt for the LLM engaged in the task (e.g., GPT-4). Given the LLM's proficiency in exploiting the data within its immediate context, the efficacy of the RAG framework is more likely constrained by the quality of the external database used, the embedding model quality, and its retrieval capabilities, rather than by the limitations of the LLM's existing knowledge base. Such a framework is referred to as the naive-RAG paradigm, and its workflow is illustrated in Fig.~\ref{fig:RAG_framework}.

The described naive-RAG paradigm was implemented using LlamaIndex \cite{Liu_LlamaIndex_2022}, a data framework designed to support LLM-based application development. The embedding model was implemented using Google Generative Language Semantic Retriever, which splits the documents into nodes and creates vector embeddings of the text of each node using \texttt{text-embedding-ada-002}.

\begin{comment}
An example of such a system is Faiss \cite{douze2024faiss}, suitable for handling extensive datasets in the order of billions of words. Faiss operates on both CPU and GPU platforms, efficiently managing vast vector databases and grouping vectors based on similarity. 
The RAG framework uses the most pertinent text chunks, identified as the top $K$ similar segments, in conjunction with the original query to accomplish its tasks

In the context of telecom knowledge, a RAG framework can be deployed on top of each release, utilizing a strong open-source model to add annotations to each document and chunk to increase the likelihood of finding the best match. Additionally, a separate model can augment the query to the RAG framework, effectively increasing the potential overlap between the given text or question and the desired chunk of text from standardization.
\end{comment}

\subsection{Performance Evaluation} 

To evaluate the efficacy of the dataset, we conducted an assessment study using a questionnaire of 100 multiple-choice questions based on 3GPP Release 15--17, focusing on series 36 and 38. These series contain specifications for radio technology for long-term evolution (LTE), LTE-Advanced, and 5G New Radio (NR). 

\subsubsection*{Automated questionnaire generation}

The questionnaire was created following a three-step procedure. First, a prompt engineering process was used to generate technical questions from the documents via the GPT-4 API. GPT-4 then categorized each question as easy, intermediate, or hard based on the following criteria:
\begin{itemize}
\item \textbf{Easy:} Questions involving fundamental concepts, widely known facts, or direct information from the documents.
\item \textbf{Intermediate:} Questions requiring understanding and application of concepts or involving moderate levels of calculations or inference.
\item \textbf{Hard:} Questions involving complex calculations, deep understanding of 3GPP standards, or integrating multiple pieces of information from different sources.
\end{itemize}

\noindent Second, the difficulty assignments were verified using the open-source LLM Mixture of Experts \cite{jiang2024mixtral} (Mistral $7\times 8\text{B}$). This resulted in 78\% of the questions being assigned the same difficulty category as GPT-4, indicating the quality of the prompt engineering process. The remaining 22\% were on the boundaries between easy-intermediate and intermediate-hard.

\noindent Third, human validation was conducted to determine the correct category for the 22\% of conflicted assignments. The questionnaire, prompts, and difficulty assignments are available open-source \cite{telcospec}. A sample question from each difficulty category is shown in the Appendix.

\subsubsection*{Performance assessment} 
\begin{figure}
\centering
\includegraphics[width=\figwidth]{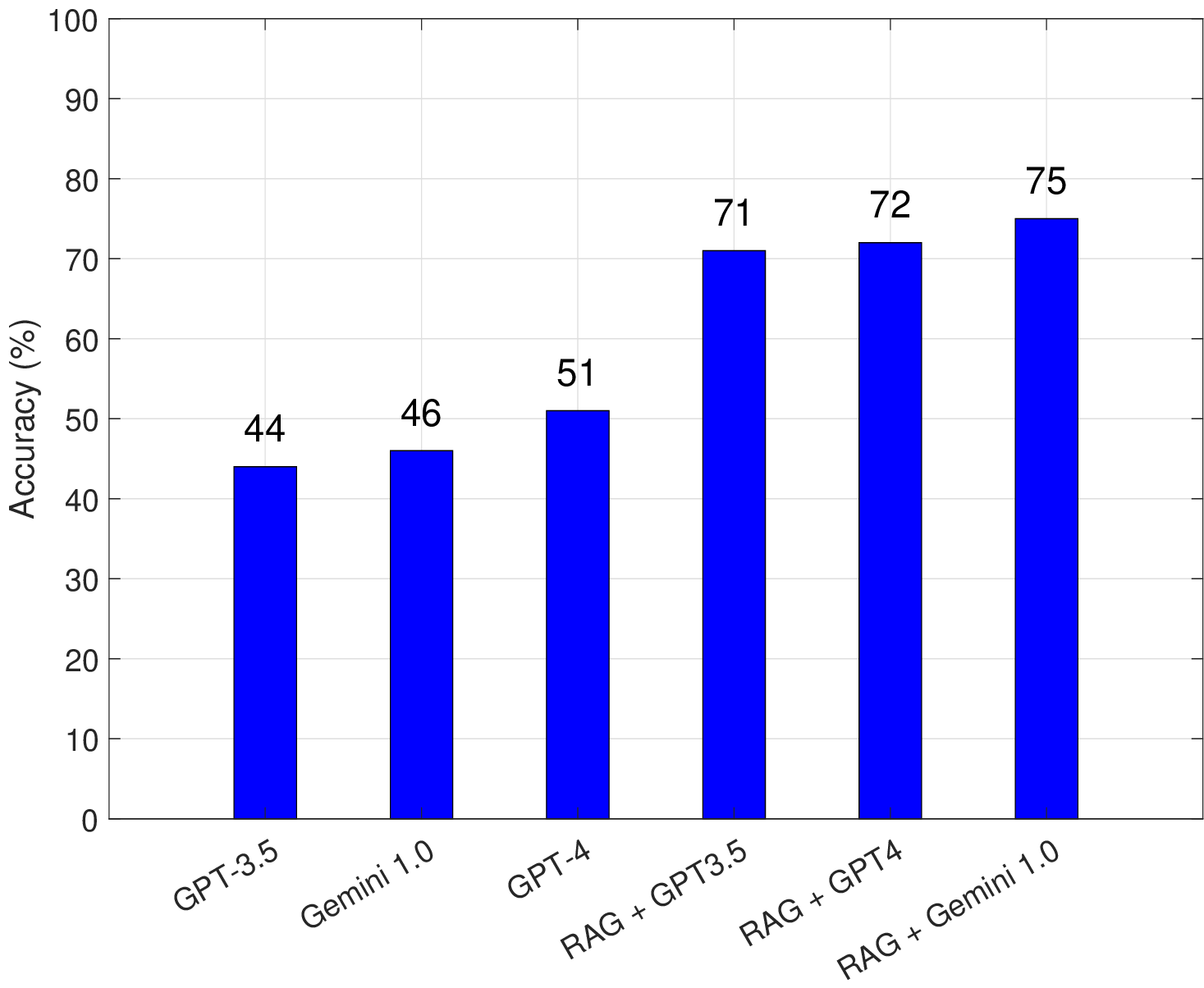}
\caption{Accuracy comparison among GPT-3.5, GPT-4, and Gemini, with and without employing naive-RAG on the TSpec-LLM dataset.}
\label{fig:Accuracy}
\end{figure}

We evaluated the performance of GPT-3.5, GPT-4, and Gemini Pro 1.0, along with the naive-RAG framework combined with each of these base models, using the created questionnaire. The chunk size and context length were kept the same for all assessed LLMs. 
Performance was measured by accuracy, defined as the percentage of questions for which the LLM selects the correct option. The results are presented in Fig.~\ref{fig:Accuracy}. All three state-of-the-art models performed similarly on domain-specific queries, with accuracies of 44\% for GPT-3.5, 46\% for Gemini Pro 1.0, and 51\% for GPT-4. These results indicate that further adaptations to the telecom domain are needed to achieve higher accuracy in responding to complex domain-specific queries. Utilizing the TSpec-LLM dataset and a simple retriever mechanism like naive-RAG boosted the accuracy to 71\%, 71\%, and 75\% for GPT-3.5, Gemini Pro 1.0, and GPT-4, respectively.

\subsubsection*{Performance by question category} 
\begin{figure}
\centering
\includegraphics[width=\figwidth]{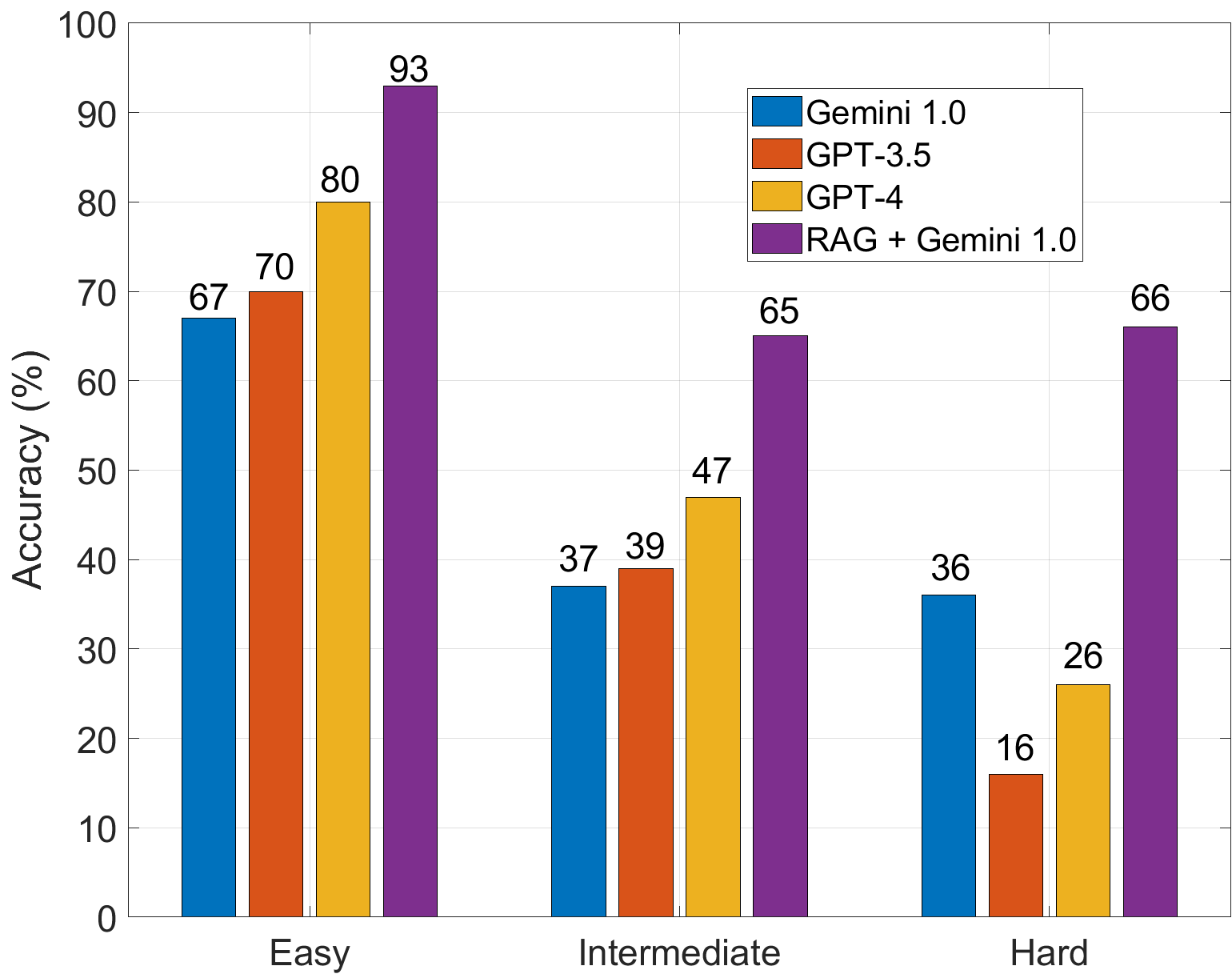}
\caption{Accuracy comparison among GPT-3.5, GPT-4, and Gemini, on specified difficulty categories, with and without employing naive-RAG on the TSpec-LLM dataset.}
\label{fig:difficulty}
\end{figure}

The accuracy of the models on each difficulty category is given in Fig.~\ref{fig:difficulty}. Since the RAG framework shows similar performance when combined with all three different base models, only the accuracy of RAG + Gemini is shown. 
GPT-4 is the base model with the highest accuracy in the easy and intermediate categories, with accuracy of 80\% and 47\%, respectively. Applying RAG on the TSpec-LLM dataset enhances this accuracy to 93\% and 65\%, respectively. 
GPT-3.5 struggles the most in the hard category, with an accuracy of 16\%, followed by GPT-4 at 26\%, and Gemini at 36\%. RAG on TSpec-LLM enhances the accuracy in this category to 66\%, providing the ability to understand the technicalities within the 3GPP specifications much better than the base models.

\subsubsection*{Confidence levels} 
\begin{figure}
\centering
\includegraphics[width=\figwidth]{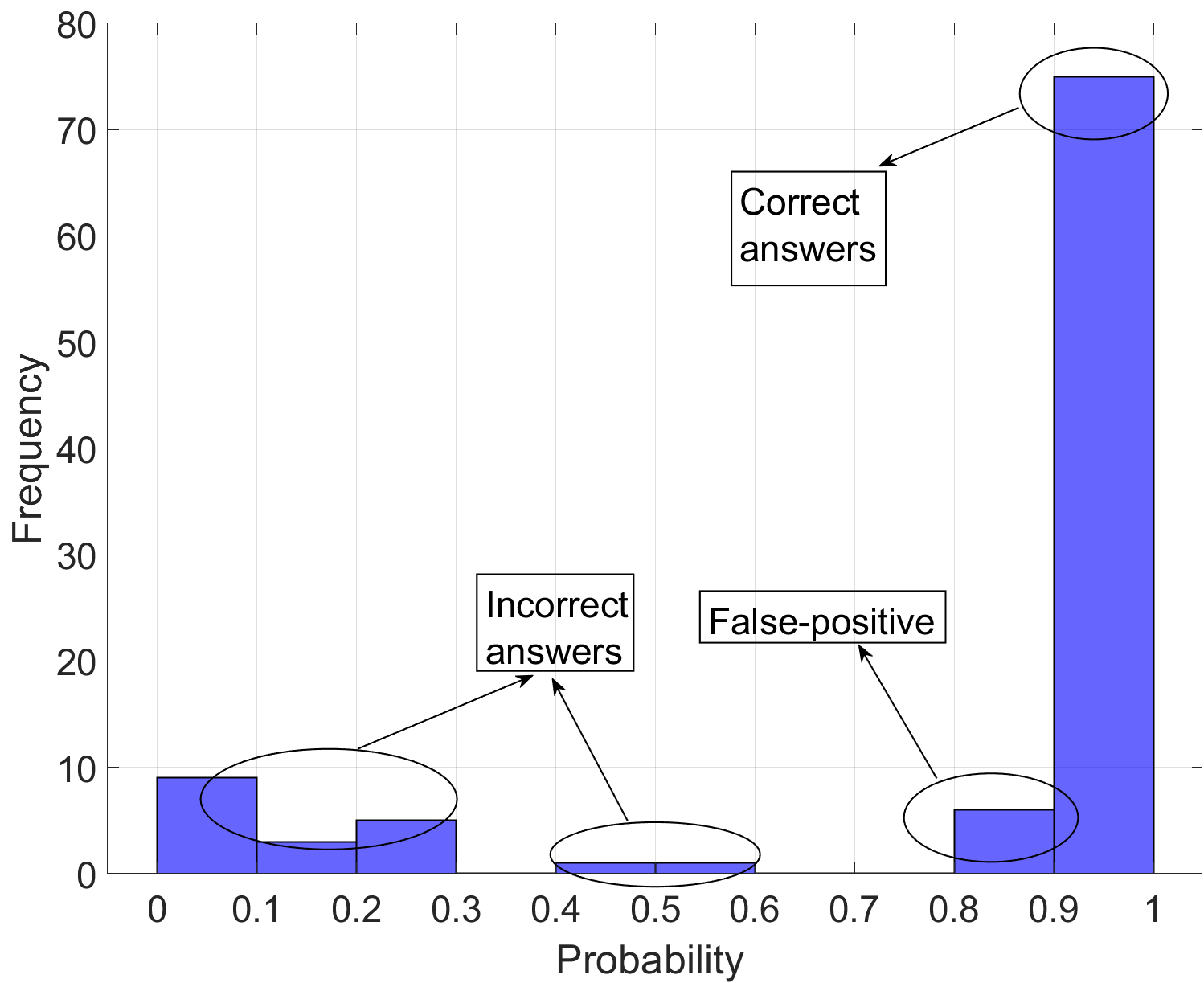}
\caption{Confidence levels, showing the frequency of the probabilities of correctness assigned to the answers generated by Gemini with RAG + TSpec-LLM.}
\label{fig:answer_prob}
\end{figure}

We analyze the confidence levels of Gemini 1.0 Pro in answering the questionnaire when applying RAG on the TSpec-LLM dataset. We do so by utilizing Google’s query engine, which for each response returns a probability indicating how confident the LLM was in answering the question based on the retrieved passages. Fig.~\ref{fig:answer_prob} shows the frequency of the probabilities assigned to the generated answers. As the overall framework achieved an accuracy of 75\%, it is observed that the model was definitively sure about the answer 72 times, with a probability of 1. The model assigned 3 answers a probability of 0.9. There were 6 false positive cases, in which the model assigned a high probability of 0.8--0.9. For the wrong answers, the model selected a lower confidence interval in 0--0.6.

\subsubsection*{RAG on different datasets} 
To better understand the effectiveness of the TSpec-LLM dataset, we conducted an analysis comparing the performance results of the naive-RAG framework using the TSpec-LLM dataset against its performance on the SPEC5G dataset \cite{karim2023spec5g}. In Table~\ref{tab:dataset_performance} we provide a breakdown of accuracy across different levels of difficulty, along with the total accuracy. The findings show that applying the naive-RAG framework to TSpec-LLM and to SPEC5G results in an overall accuracy of 75\% and 60\%, respectively. The advantage provided by TSpec-LLM can be attributed to its retention of the original content from tables and formulas, its inclusion of a complete collection of all 3GPP documents, and the preservation of their original structure without sampling or consolidation into a single file.

\begin{table}[h]
\centering
\begin{tabular}{@{}lcccc@{}}
\toprule
\textbf{Method}             & \textbf{Easy} & \textbf{Intermediate} & \textbf{Hard} & \textbf{Overall} \\ 
\midrule
RAG + TSpec-LLM          & 93\%          & 65\%            & 68\%          & 75\%             \\
RAG + SPECSG         & 83\%          & 47\%            & 57\%          & 60\%             \\
\bottomrule
\end{tabular}
\caption{Performance achieved when applying the RAG framework to the TSpec-LLM and SPEC5G datasets, respectively.}
\label{tab:dataset_performance}
\end{table}

\section{Conclusions and Future Directions}

In this paper, we introduced \emph{TSpec-LLM}, a comprehensive open-source 3GPP dataset tailored for research utilizing LLMs. The dataset includes all 3GPP specification documents from 1999 to 2023, covering Releases 8 to 19. We detailed the dataset creation process and evaluated state-of-the-art LLMs (GPT-3.5, GPT-4, and Gemini) using a questionnaire focused on radio technology specifications defined by 3GPP. To assess the dataset's quality, we employed a RAG framework to retrieve relevant information before answering questions. Our results showed that combining TSpec-LLM with a naive-RAG framework enhances GPT-3.5, Gemini 1.0 Pro, and GPT-4 overall accuracy from 44\%, 46\%, and 51\% to 71\%, 75\%, and 72\%, respectively. In particular, naive-RAG on TSpec-LLM boosts the accuracy on the most difficult questions (requiring a deep understanding of 3GPP standards) from 16--36\% to 66\%.

While showing promising results, the proposed naive-RAG framework does not follow an optimized indexing structure, leading to lower retrieval quality and inaccurate responses at times. This can occur when not all nodes within the retrieval set correlate with the query, or due to low recall, where not all relevant nodes are retrieved, preventing the LLM from obtaining sufficient context to synthesize an answer. To address these indexing issues, future work will focus on advanced-RAG frameworks capable of optimizing indexing through methods such as sliding windows, fine-grained segmentation, and metadata augmentation. 
%Specifically, we will focus on augmenting both the query and the text chunk using a capable open-source model such as LLaMA-3 70B, effectively increasing the overlap between the query and the desired text chunk in the embedding space. 

Additionally, future work will involve developing a more comprehensive questionnaire for a specific task (e.g., answering release- or series-specific queries), following the generation steps outlined in this paper. Such questionnaire could be integrated with the corresponding portion of the TSpec-LLM dataset. The integrated dataset would serve as a knowledge base for fine-tuning smaller open-source language models, such as Phi3 \cite{abdin2024phi3}, to construct a robust telecom-specific model. This model could then be deployed offline using tools like Ollama \cite{ollama}, and it would be capable of operating locally in a browser environment through libraries like WebLLM \cite{web-llmM}.

%\appendices
\section*{Appendix A: Samples from the Questionnaire}
\label{questionnaire}

\textbf{Question 1:} What is the maximum directional gain of an antenna element?
\begin{enumerate}
    \item 5 dBi.
    \item 8 dBi.
    \item 10 dBi.
    \item 12 dBi.
\end{enumerate}
\textbf{Correct answer:} 2) 8 dBi.\\
\textbf{Explanation:} In Table 7.3-1 under the antenna modelling section, the maximum directional gain of an antenna element is specified as 8 dBi.\\
\textbf{Category:} 3GPP TR 38.901 V16.1.0\\
\textbf{Difficulty:} Easy\\

\textbf{Question 2:} What is the adjacent beam spacing computation based on in a single satellite simulation?
\begin{enumerate}
    \item Beam diameter.
    \item Beam height.
    \item 3dB beam width of the satellite antenna pattern.
    \item Satellite altitude.
\end{enumerate}
\textbf{Correct answer:} 3) 3dB beam width of the satellite antenna pattern.\\
\textbf{Explanation:} As per Table 6.1.1.1 Adjacent beam spacing.\\
\textbf{Category:} 3GPP TR 38.821 V16.2.0\\
\textbf{Difficulty:} Intermediate\\

\textbf{Question 3:} For RMa-AV and UMa-AV in LOS conditions, what are the desired angular spreads (ASA and ASD)??
\begin{enumerate}
    \item 0.2 for both ASA and ASD.
    \item 0.1 for both ASA and ASD.
    \item 0.2 for ASA and 0.1 for ASD.
    \item 0.1 for ASA and 0.2 for ASD.
\end{enumerate}
\textbf{Correct answer:} 1) 0.2 for both ASA and ASD.\\
\textbf{Explanation:} 3GPP TR 36.777 Table B.1.1-1 specifies that for RMa-AV LOS, the desired angular spreads (ASA and ASD) are 0.2 each.\\
\textbf{Category:} 3GPP TR 36.777\\
\textbf{Difficulty:} Hard

\section*{Appendix B: TSpec-LLM Dataset}
\label{dataset_vis}

\begin{figure}
\centering
\includegraphics[width=.99\figwidth]{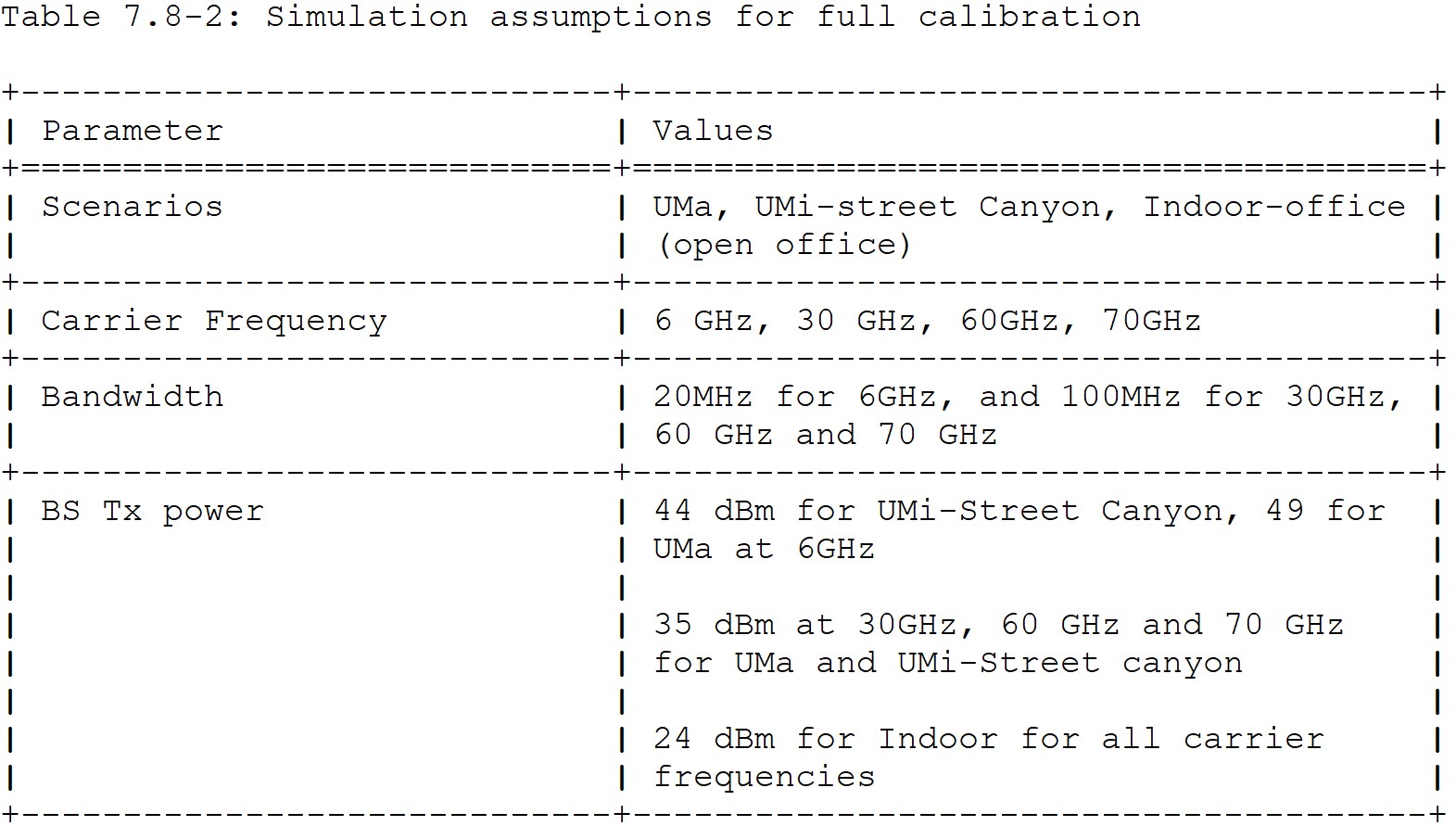}
\caption{Example of the content in the TSpec-LLM dataset, showing the simulation parameters in \cite[Table~7.8-2]{3GPP38901}.}
\label{fig:dataset_vis_tab}
\end{figure}

The TSpec-LLM dataset is hosted on Hugging Face \cite{telcospec} and it can be downloaded by following the installation guide provided. Its total size is approximately 15\,GB. In TSpec-LLM, specifications are categorized by releases, followed by their corresponding series. For each series, the original files are preserved in .docx format, along with the processed files in  markdown (.md) format, which can be used for conducting research using LLMs. Fig.~\ref{fig:dataset_vis_tab} shows an example of TSpec-LLM, where the content of Table~7.8-2 from \cite{3GPP38901} is preserved within the markdown files. This type of content is important as it contains system parameters, configurations, and other technical details. Fig.~\ref{fig:dataset_vis_eqn} shows a comparison between an equation represented in its original .docx format and its corresponding representation in TSpec-LLM, using LaTex format which facilitates processing by state-of-the-art LLMs.

{\begin{figure}
    \centering
    \subfloat[Original Equation~7.6-9 from \cite{3GPP38901} (.docx).]{
    \includegraphics[width=0.99\figwidth]{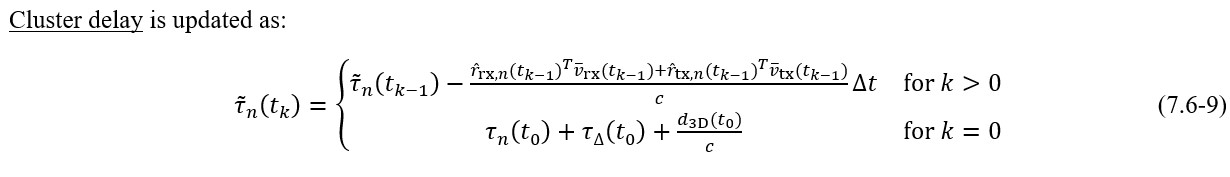}
    }\\
    \hspace{3mm}
    \subfloat[Equation representation in TSpec-LLM (.md).]{
        \includegraphics[width=0.99\figwidth]{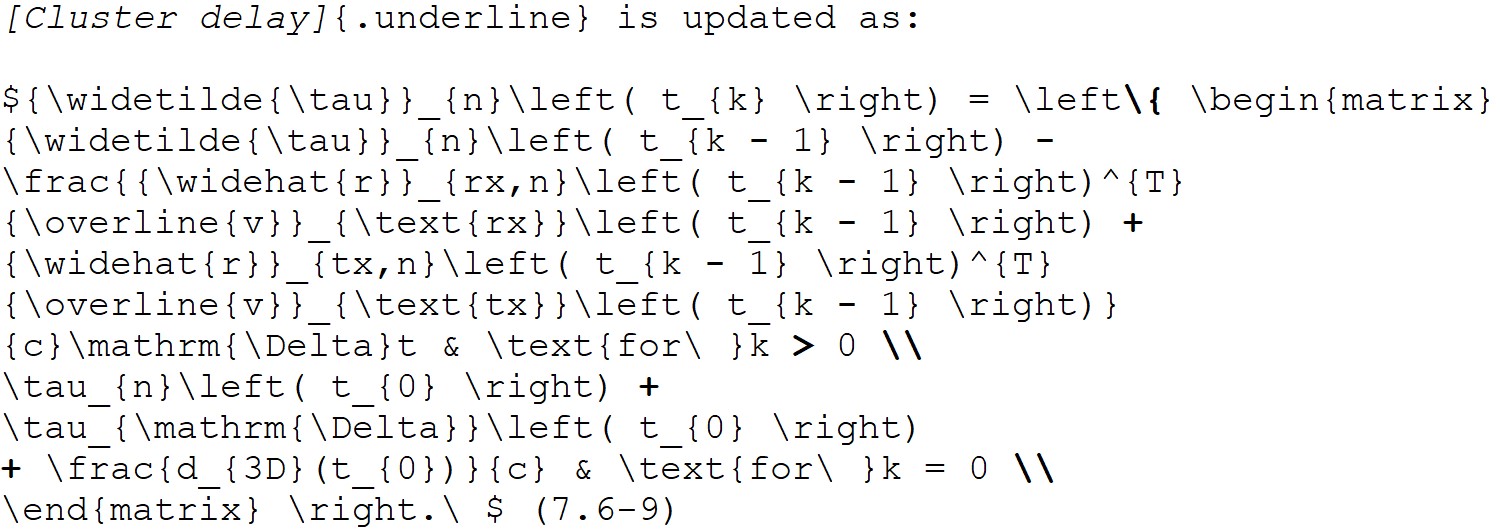}
    }
    \caption{Example of the content in the TSpec-LLM dataset, showing: (a) an equation in the original document (.docx) \cite[Eqn.~7.6-9]{3GPP38901} and (b) the corresponding representation (.md) in TSpec-LLM.}
    \label{fig:dataset_vis_eqn}
\end{figure}}

\begin{comment}
\textbf{Step 1:} 
Make sure you have Git LFS installed (\url{https://git-lfs.com}) 
\begin{lstlisting}
git lfs install
\end{lstlisting}

\textbf{Step 2:} 
Clone the dataset repository 
\begin{lstlisting}
git clone https://huggingface.co/datasets/rasoul-nikbakht/TSpec-LLM
\end{lstlisting}
\end{comment}

%\newpage
\bibliographystyle{IEEEtran}
\bibliography{journalAbbreviations, main}

\end{document}